# Could planet/sun conjunctions be used to predict large (>=Mw7) earthquake?

Pierre Romanet, Earthquake and Tsunami Research Division, NIED, Tsukuba, Ibaraki, Japan

## Abstract

No.

## Introduction

Following the recent Mw 7.8 Kahramanmaraş, Türkiye earthquake sequence on 6 February 2023, the assertion that planet/sun alignments and lunar phases may help to predict earthquake became widespread in some bad quality news and social medias. In the following, we will call this alignment of three celestial bodies a conjunction, although the correct word must be a syzygy.

Usually, this assertion is promoted by choosing carefully period of time over which it occurs and showing specific earthquakes at which it occurs. Also they usually do not mention that these events do happen extremely frequently, and that most of the time, these alignments are not followed by significant earthquakes.

The only literature available about it put into question fundamental physics without any proof (Omerbashich, 2011; Safronov 2022), or does not show the background rate of conjunctions (Awadh, 2021).

The major logical flaw in their analysis is showing only events that are working while not paying attention to the total quantity of conjunctions (see Khalisi, 2021, Zanette 2011). Indeed, if conjunctions are very common, it is easy to associate them with earthquakes.

This assertion can be seen as a more evolved version of that the moon phase is changing the earthquake. The moon phase theory, has been debated for a long time by seismologists (Schuster, 1897), and the question is still not completely answered yet (Ide et al., 2016; Hough, 2018; Kossobokov and Panza 2020; Zaccagnino et al., 2022). In some regions, slow-earthquakes like tremors (Nakata et al 2008, Rubinstein et al., 2008), or low frequency earthquake (Thomas et al., 2012) are influenced by tides. Depending on the area, the time in the seismic cycle (Tanaka 2010, 2012; Peng et al., 2021) and the focal mechanism of the earthquakes (Tsuruoka et al., 1995), it may have some influence or not. Overall, it seems to have an influence (Yan et al,. 2023), at least for some regions/period or time, that may be incorporated in long term probabilistic earthquake forecasting (Ide et al., 2016). Rigorous attempt to perform short term prediction with the idea that before a large earthquake, smaller earthquakes would be more tide-sensitive as the crust is approaching critical strength, was proven to be ineffective (Hirose et al., 2022).

While for the moon/earth/sun alignements, there exists a physical mechanism by which the stresses are changing in the crust (the gravity), and therefore may weakly influence earthquake occurence (Ide et al., 2016), there is no such mechanism for planets/sun alignements, because the electromagnetic and gravity fields by celestial body other than sun and moon are usually extremely small when they reach the Earth. Therefore, invoking "electrodynamic", "resonance", and "molecule" as if they were keywords to explain the phenomena leading to this assertion only reflects the lack of scientific knowledge of the persons promoting this theory.

In this paper, we are testing the planet/sun alignment, together with the moon phase systematically over a 69 years period of time using global catalog of earthquakes. We are systematically comparing the percentage of earthquakes linked with conjunction(s) with the percentage of the time that conjunction(s) are happening. This assertion that planet/sun alignment is promoting earthquakes would be valid only if it is happening more frequently than conjunctions themselves. We also calculated significance of our results, by calculating the p-value, making the null hypothesis that earthquakes follow a binomial distribution during the period with the probability given by the probability of conjunctions.

## Method

We first chose the ISC-GEM catalog (Storchak et al., 2013; 2015; Di Giacomo et al., 2018) and selected earthquakes of Mw>7 over the period 1950/01/01-2018/12/31. The reason of selecting the year 1950, is because the catalog starts to be complete for shallow events (>60km) and for Mw>=7 at years 1918–1939 (Michael, 2014). We chose the 10 years delay as a margin to be sure not to miss Mw 7 earthquakes which may flaw the analysis.

To calculate each planet/sun alignment, we took advantage of the Astropy package in python (The astropy collaboration et al., 2018, 2022), that allows to calculate the position of any planets in the solar system, the sun, and the moon at any time. For each day covering the period of the earthquake catalog, we calculated if there was a conjunction or not. We used the NASA JPL ephemeris model "DE430". We did not take into account leap seconds in the calculation of the day, because the offset is less than a minute for the considered period.

The celestial body included are: the Sun, Mercury, Venus, the Earth, Mars, Jupiter, Saturn, Uranus, and Neptune.

For each triplet of given three celestial bodies A, B and C in the solar system, we calculated their positions in International Celestial Reference System (ICRS).

We then calculated each vector $\overrightarrow{AB}$, $\overrightarrow{BC}$ and $\overrightarrow{AC}$ and the associated norms $||\overrightarrow{AB}||$, $||\overrightarrow{BC}||$ and $||\overrightarrow{AC}||$. The vector that has the longest norm shows the two bodies whose distance is the greatest, hence we can find the body that is in the middle. For example, if $||\overrightarrow{AC}||$ is the greatest distance, we can guess that the celestial body B is in the middle. Finally, we can calculate the angle between $\overrightarrow{AB}$ and $\overrightarrow{BC}$ as:

$$\theta = \frac{180}{\pi} \arccos\left(\frac{\overrightarrow{AB} \cdot \overrightarrow{BC}}{|\overrightarrow{AB}||\overrightarrow{BC}|}\right) \text{ in degree}$$

When the angle $\theta$ was smaller than a threshold $\theta_{thr}$ we set that there was an alignment of the celestial bodies for the day.

For the moon phase, we calculated the projection of the moon on the ecliptic plane (the plane that contains the orbital of the Earth). Then, we try to find if the projection on this plane was in opposition (full Moon) or in conjunction (new Moon) with the sun from the Earth. A threshold of 6.5° was used, this threshold is chosen because the average orbital of the moon around the Earth during one day is around 12°.

# Results

| Number of days associated with conjunction(s) | Number of earthquakes associated with conjunction(s) | Number of days associated with full/new moon | Number of earthquakes associated with full/new moon | Number of days associated with both full/new moon and conjunction(s) | Number of earthquakes associated with both full/new moon and conjunction(s) |
|---|---|---|---|---|---|
| 19565/25202 (77.63%) | 640/813 (78.72%) $p_{value} = 0.23$ | 1743/25202 (6.92%) | 58/813 (7.13%) $p_{value} = 0.40$ | 1349/25202 (5.35%) | 52/813 (6.40%) $p_{value} = 0.09$ |

Table 1: comparison of the frequency of a particular event (for example a conjunction), and the frequency of an earthquake that can be associated to the event during the period 1950/01/01-2018/12/31. The threshold used here to define a conjunction is $\theta_{thr} = 3°$. The calculated p-value is one-sided, for the null hypothesis that the earthquakes follow a binomial law with the probability given by the frequency calculated with the number of days.

The results are presented in the above chart (table 1). The total period consists of 25202 days, among which 19565 days are associated with conjunctions. So that 78% of the time, there is at least one conjunction on the day. For the same period, there are 813 earthquakes, among which 640 are associated with conjunctions, so that 79% percent of earthquakes are associated with conjunctions.

We did the same study for earthquakes associated with full or new moon, as well as for earthquakes associated with both full or new moon, and at least one conjunction. The percentage of days associated with either full or new moon is 7% (1743/25202), very much the same as the number of earthquakes that happened during full or new moon 7% (58/813). Finally, there are 5% (1349/25202) of days, and 6% (52/813) of earthquakes associated with both full or new moon, and at least one conjunction.

We can formulate the null hypothesis that earthquakes follow a binomial law with the probability $p$ given by the number of days that are associated with conjunctions:

$P[k|n|p] = \binom{n}{k} p^k (1-p)^n$, where $P$ is the probability to observe $k$ earthquakes that are associated with at least one conjunction in the total number of earthquakes $n$. Because $n$ is large in our sample, we can approximate the binomial distribution by a normal law:

$$P[k|n|p] \simeq \frac{e^{-\frac{1}{2}\left(\frac{k-np}{\sqrt{np(1-p)}}\right)^2}}{\sqrt{2\pi np(1-p)}},$$

finally the single-side p-value will be:

$$p_{value} = \frac{1}{2} - \frac{1}{2}\text{erf}\left(\frac{k - np}{\sqrt{2np(1-p)}}\right), \text{ if } k > np$$

The p-value represents the probability to obtain a value worst or equal than the calculated value. Usually, a value $p_{value} < 0.05$ would mean that we can reject the null hypothesis, meaning that there is only 5% chance that we get such a bad result.

We choose a single-side p-value because this will favor the rejection of the null hypothesis (the single side p-value is lower than the two-side p-value), hence it is siding with the hypothesis that conjunctions are linked to earthquakes. Given that the p-value for earthquakes associated with conjunction(s) is 0.23, we cannot reject the null hypothesis, hence the difference between the observed value of earthquake linked with conjunction and the total number of conjunction is not significant. The same analysis can be done for earthquakes associated with full/new moon, or earthquakes that are associated with both full/new moon and at least one conjunction. In these two cases, the p-value is also high enough ($p_{value} > 0.05$) so that we cannot reject the null hypothesis.

# Discussion and Conclusion

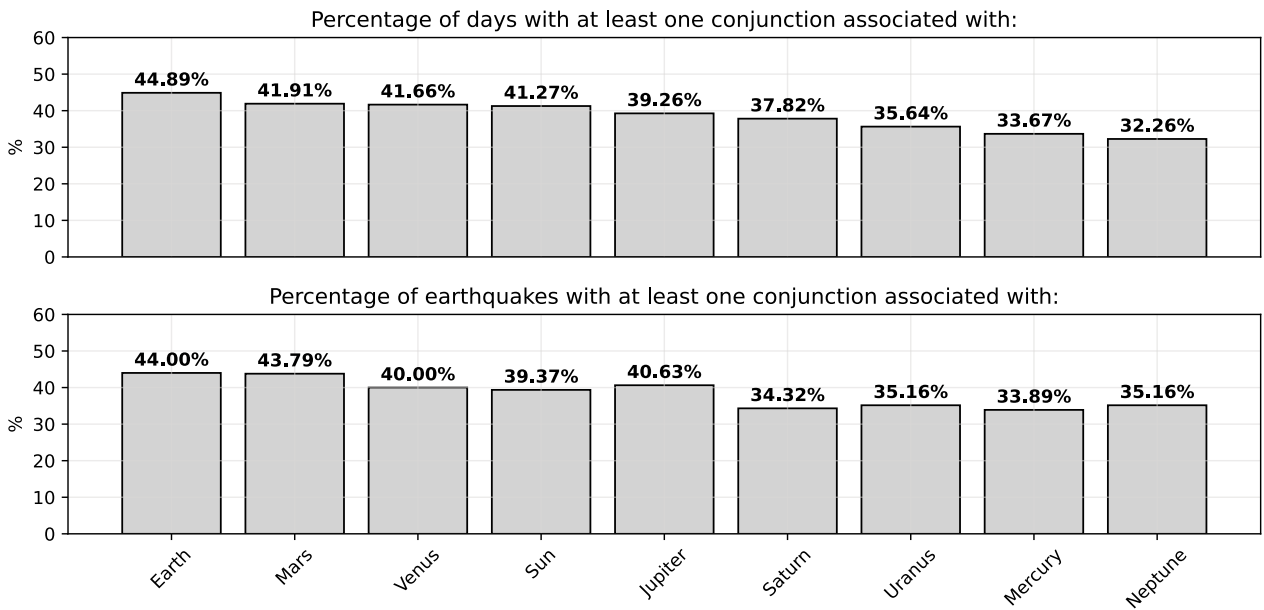

Figure 1: Comparison of the percentage of days involving at least one conjunction associated with a given planet, and the percentage of earthquakes linked with at least one conjunction associated with a given planet. The threshold angle to define a conjunction is $\theta_{thr} = 3°$.

The frequency of earthquakes associated with conjunction(s) and the frequency of conjunctions are pretty much the same, and the difference is statistically non-significant (all the p-values are larger than 5%). It means that we cannot reject the hypothesis that earthquake are occurring following a binomial law given the time period.

The fact that the null hypothesis cannot be rejected does not mean either that this is the true hypothesis. It is known that earthquakes are not completely random, especially because of aftershocks, and aftershocks have not been removed here. It just means that given this earthquake catalog, we cannot decipher to reject it.

Nether-the-less, the assertion that earthquakes are linked with conjunction is unlikely based on our results. For such a strong claim, that earthquakes can be predicted using conjunctions and moon phase, because it would have extremely important societal outcome, it would need very significant results and hence associated with very law p-value. This is far from being the case here.

We also tried to find if a planet was more often that others associated with conjunctions (figure 1). It seems not the case because the difference between the percentage of planet/sun involved at least in one conjunction during one day is within 3% the same as the percentage of earthquakes that can be at least associated with a given planet/sun in a conjunction.

Finally, we tried to see if a conjunction was more often than others associated with earthquake occurence (figure 2). The results are less clear, because for a given conjunction, the percentage of one particular conjunction during the whole period is small (<2% for the conjunction that is the most frequent), so that the number of earthquakes sampling this conjunction is also very small. This leads to a large variability. However, we can still say that the overall trend is respected, the conjunctions that are the most frequent are most often associated with earthquakes.

The change of the threshold for conjunction does not change the results, and the same conclusion can be made. If the threshold angle is too small, we may miss some conjunctions because the orbital plane is not exactly the same for each planet. For example, the results with the threshold of 2% is given in appendix (table 2). Reducing the threshold angle mainly reduces the percentage of time conjunctions are happening and reduces in the same way the percentage of earthquakes that are associated with conjunctions.

Persons defending the assertion of planetary/sun conjunctions may continue arguing that I still did not look at a particular association of conjunctions, or association with only full moon. This is true. But given the number of possible associations, it is impossible to test them all. If so, they are very welcomed to indicate these specific associations, so that it can be tested rigorously and scientifically, keeping in mind that normally the person making assertions should be the one proving them.

The alignment of three planets/sun is actually something extremely ordinary in the solar system that is happening close to everyday (for the threshold $3°$, it happens 78% of the time). Finding a syzygy on the day of an earthquake is therefore normal, moreover if we start looking at some days before and after an earthquake. We showed that the percentage of earthquake associated with at least one conjunction is actually very similar to the percentage of the time where there is at least a conjunction, and that the difference between the two is not statistically significant. Hence, there is no significant effect of planet/sun alignment or moon effect on the occurence of large earthquakes, and it can certainly not be used to provide short term prediction of earthquakes. Finally to plagiarize Khalisi, 2021, "Sooner or later there will be another earthquake close to a ***conjunction***, and the self-proclaimed prophets will have their joy."

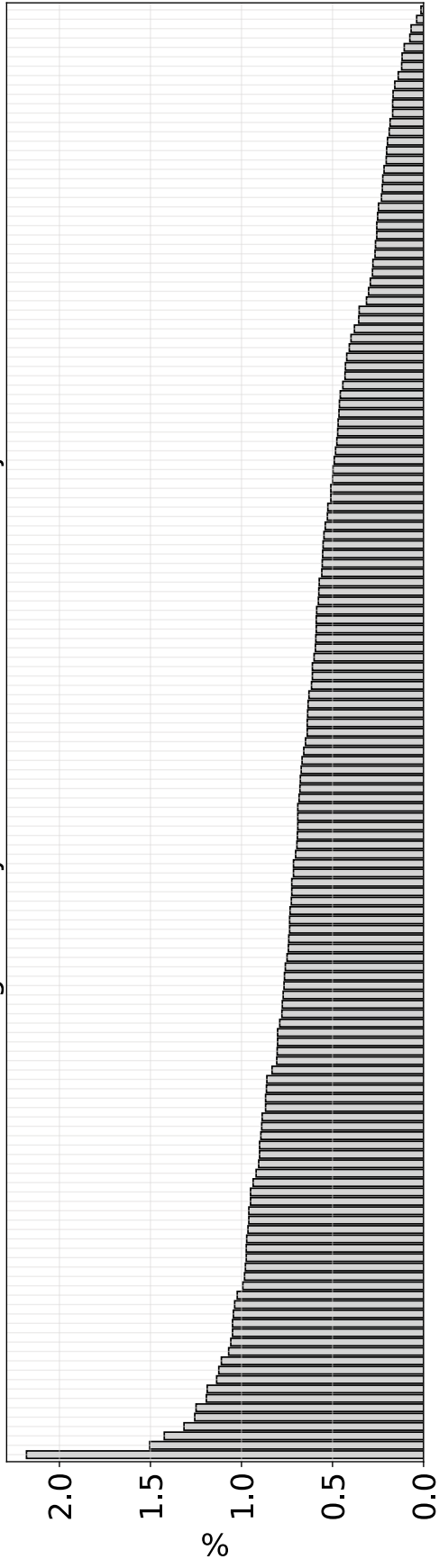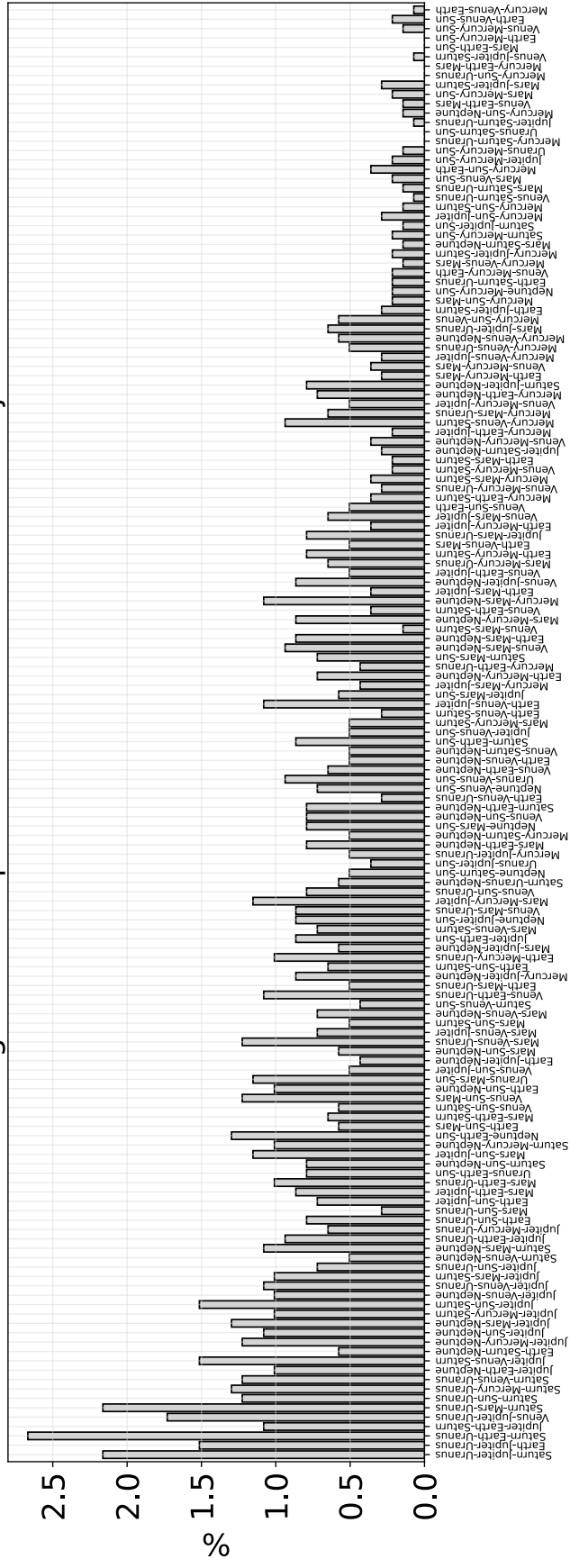

Figure 2: Comparison of the percentage of days that a specific conjunction happen, with the percentage of earthquakes that can be linked with the same specific conjunction. The threshold angle to define a conjunction is $\theta_{thr} = 3°$.

## Acknowledgement

I would like to thank all the seismologists/geologists/scientists that supported me to write this article. Special thanks to Martijn van den Ende and Sylvain Barbot, because I would have never thought about publishing a paper on this topic. I would also like to thank Susan Hough that allowed me to plagiarize her abstract.

## Data

International Seismological Centre (2018), ISC-GEM Earthquake Catalogue, https://doi.org/10.31905/d808b825

# Appendix

| Number of days associated with conjunction(s) | Number of earthquakes associated with conjunction(s) | Number of earthquakes associated with full/new moon | Number of earthquakes associated with full/new moon | Number of days associated with both full/new moon and conjunction(s) | Number of earthquakes associated with both full/new moon and conjunction(s) |
|---|---|---|---|---|---|
| 13908/25202 (55.19%) | 463/813 (56.95%) $p_{value} = 0.16$ | 1743/25202 (6.92%) | 58/813 (7.13%) $p_{value} = 0.40$ | 976/25202 (3.87%) | 34/813 (4.18%) $p_{value} = 0.32$ |

Table 2: comparison of the frequency of a particular event (for example a conjunction), and the frequency of an earthquake that can be associated to the event during the period 1950/01/01-2018/12/31. The threshold used here to define a conjunction is $\theta_{thr} = 2°$. The calculated p-value is one-sided, for the null hypothesis that the earthquakes follow a binomial law with the probability given by the frequency calculated with the number of days.